# BioSEAL: In-Memory Biological Sequence Alignment Accelerator for Large-Scale Genomic Data

Roman Kaplan, Leonid Yavits and Ran Ginosar

**Abstract**—Genome sequences contain hundreds of millions of DNA base pairs. Finding the degree of similarity between two genomes requires executing a compute-intensive dynamic programming algorithm, such as Smith-Waterman. Traditional von Neumann architectures have limited parallelism and cannot provide an efficient solution for large-scale genomic data. Approximate heuristic methods (e.g. BLAST) are commonly used. However, they are suboptimal and still compute-intensive.

In this work, we present BioSEAL, a Biological SEquence ALignment accelerator. BioSEAL is a massively parallel non-von Neumann processing-in-memory architecture for large-scale DNA and protein sequence alignment. BioSEAL is based on resistive content addressable memory, capable of energy-efficient and high-performance associative processing.

We present an associative processing algorithm for entire database sequence alignment on BioSEAL and compare its performance and power consumption with state-of-art solutions. We show that BioSEAL can achieve up to 57× speedup and 156× better energy efficiency, compared with existing solutions for genome sequence alignment and protein sequence database search.

──────────── ◆ ────────────

## 1 INTRODUCTION

Improvement in *genome sequencing* technology has led to a reduction in the cost of sequencing and an increase in genomic database sizes, outpacing Moore's law. An estimated 100 million to 2 billion human genomes could be sequenced by 2025, surpassing other big data aggregators such as YouTube and Twitter [1]. Moreover, it is soon expected that the cost of sequencing a genome will drop below a hundred dollars [2], enabling population-scale genomic datasets. These datasets can enable improved identification of genetic correlation between complex traits and diseases [3]. Genomic data is already being used today for the detection of Parkinson's [4] and Alzheimer's [5] diseases, and for better treatment of cancer [6].

Large genomic datasets can be analyzed in different ways; each provides different insights and most require sequence alignment [1]. Sequence alignment is a fundamental problem in genomics. It aims to find how one sequence could transform to the other by using scores for each of the possible character transformations: insertion, deletion and gap.

Whole genome alignment is a form of population genomic data analysis, used for genome annotation [7] and phylogeny reconstruction [8][9]. A single whole genome alignment between human and mouse consumes ~100 CPU hours [10]. By 2025, ~2.5 million species genomes are expected to be available, requiring roughly 50-100 trillion such whole genome alignments, six orders of magnitude more than is possible today in reasonable time. Protein databases also use alignment to determine relatedness between two genes or find the function of a gene within an organism's genome [11].

The Smith-Waterman (S-W) [12] local sequence alignment algorithm provides an optimal solution for comparing two biological sequences (protein or DNA). However, it has a high computational complexity of score calculations, $O(n \cdot m)$, where $n$ and $m$ are the lengths of the sequences being compared. Therefore, this algorithm is not commonly used for homology detection [13], where multiple sequences must be aligned, for phylogeny tree construction [13], or when searching a single sequence against a sequence database [11][14]. Therefore, common methods for all these tasks [15][16] are based on heuristic search to reduce the computational complexity to $O(\max(n,m))$. However, these methods are suboptimal in the sense that they may miss the highest similarity sequence [17].

This work presents BioSEAL, a massively parallel processing-in-memory Biological SEquence ALignment accelerator. Its main applications include (1) pairwise alignment of long, whole-genome, DNA sequences and (2) alignment of a query sequence with an entire database of sequences, protein or DNA, so that the highest score is always attached to the highest similarity sequence. This score indicates the functionality of a protein [18], the distance between organisms in a phylogenetic tree [8], or the role of a gene [7].

BioSEAL is based on a novel NAND Resistive Content Addressable Memory (ReCAM). BioSEAL simultaneously functions as data storage and as a massively parallel SIMD accelerator that performs *in-situ* associative computations, overcoming the von Neumann bandwidth bottleneck. The novel architecture results in increased performance and reduced energy consumption. We have further modified the ReCAM to speed up associative operations. The modification exploits repeating write values and uses a single write cycle, instead of several, to write the same value in multiple locations. We call this approach *batch-write*.

Previous work [19] presented PRINS, a NOR crossbar array architecture, using non-batch write associative processing to perform only pairwise sequence alignment.

- Roman Kaplan, E-mail: romankap@gmail.com.
- Leonid Yavits, E-mail: yavits@technion.ac.il.
- Ran Ginosar, E-mail: ran@ee.technion.ac.il.

*Authors are with the Department of Electrical Engineering, Technion-Israel Institute of Technology, Haifa 3200000, Israel.*

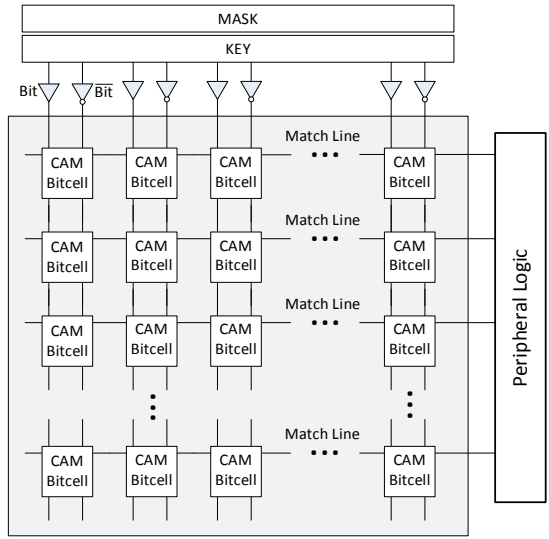

Figure 1: Content addressable memory array, composed of bitcells, registers (MASK and KEY, top), and peripheral logic (right).

This work makes the following contributions:
- A novel NAND-based ReCAM crossbar array
- Batch-write associative processing in ReCAM
- BioSEAL, an energy-efficient processing-in-memory accelerator architecture for biological sequence alignment
- Permanent storage for whole-genome and protein databases, enabling *in-situ* processing without the need for voluminous data transfers
- An associative processing algorithm for whole-database sequence alignment

We compare the performance and energy efficiency of BioSEAL with other solutions and show that BioSEAL outperforms them by up to 57× with up to 156× better energy efficiency

The paper is organized as follows. Section 2 provides background on content addressable memories, associative processing, and sequence alignment algorithms. Section 3 explains NAND ReCAM. Section 4 discusses batch-write. The BioSEAL architecture is described in Section 5. Section 6 shows how the pairwise sequence is mapped onto BioSEAL and demonstrates an entire database sequence alignment algorithm on BioSEAL. Performance and energy efficiency evaluations are the focus of Section 7. Related work is discussed in Section 8 and conclusions are presented in Section 9.

## 2 BACKGROUND

The following subsections describe the basic concepts that form the basis for this work. The first subsection covers the associative processing compute model and its enabling hardware, the content addressable memory. The second covers the Smith-Waterman local sequence alignment algorithm, demonstrates its compute requirements, and covers global and semi-global sequence alignment algorithms.

### 2.1 Content Addressable Memory and Associative Processing

An associative processor (AP) is a non-von-Neumann in-memory computer [20]. The AP is based on a content addressable memory (CAM), which allows comparing the entire dataset to a search pattern (key), tagging the matching row, and writing another pattern to all tagged rows. The architecture of a content addressable memory that supports associative processing is presented in Figure 1. The memory array consists of bitcells, which may contain resistive elements [20]. The memory array is organized in bit-columns and word-rows. A word-row makes a Processing Unit (PU). Several special registers are appended to the memory array. The KEY register contains a key data word to be written or compared against. The MASK register defines the active fields for write and read operations, enabling bit selectivity. The TAG register marks the rows that are matched by a compare operation and may be affected by a write operation.

AP does not perform computations in the conventional sense. Most arithmetic and logic operations can be structured as series of Boolean functions, implemented by the AP as follows. The dataset is stored in CAM, typically one data element per CAM row (constituting a Processing Unit, PU). The AP controller sequentially matches all possible *input* combinations of a function's arguments against the entire CAM content. The matching CAM rows are tagged, and the corresponding function values (precalculated and embedded in AP microcode) are written into the designated *output* fields of the tagged rows.

For an $m$-bit argument $x$ ($x \in$ dataset), any Boolean function $b(x)$ has at most $2^m$ possible values ("at most" because of 'don't cares'). Therefore, a brute-force approach would incur up to $O(2^m)$ cycles, regardless of the dataset size.

More efficiently, arithmetic operations can be performed in AP in a word-parallel, bit-serial manner, reducing time complexity from $O(2^m)$ to $O(m)$. For instance, vector addition may be performed as follows [21]. Suppose that two $m$-bit CAM columns hold vectors A and B; the sum S=A+B is written onto another $m$-bit column S (Figure 2). A one-bit column C holds the carry bit. The operation is carried out as $m$ single-bit additions (1):

$$C[:] \mid S = A[:]_i + B[:]_i + C[:], \quad i = 0,\ldots,m-1 \quad (1)$$

where $i$ is the bit index, and ':' means all elements of the vector, and $C$ and $S$ are, respectively, the carry and sum bits. The single-bit addition is carried out in a series of steps. In each step, one entry of the full adder truth table (a three-bit input pattern, Figure 2a) is matched against the contents of the $A[:]_i$, $B[:]_i$, $C[:]$ bit columns and the matching rows (PUs) are tagged; the logic result (two-bit output of the truth table, Figure 2a) is written into the $C[:]$ and $S_i[:]$ bits of all tagged rows. During that operation, all but five (three input and two output) CAM bit-columns are masked out in each step. Overall, eight steps of one compare and one write operation are performed to complete a single-bit addition over all CAM rows (i.e., over all vector elements), regardless of the lengths of vectors A and B.

A snapshot of such vector addition, for $m = 4$, zero bit of the vector elements, and the 2nd entry of the truth table is shown in Figure 2. Figure 2a shows the truth table with the 2nd entry marked out. Figure 2b and c show, respectively, compare and write operations, against the backdrop of the CAM map. One 4-bit input vector occupies bit-columns 0-3 and another occupies bit-columns 4-7. The 4-bit output vector S is reserved columns in 8-11, while bit column 12 is used for storing and updating the carry bit C.

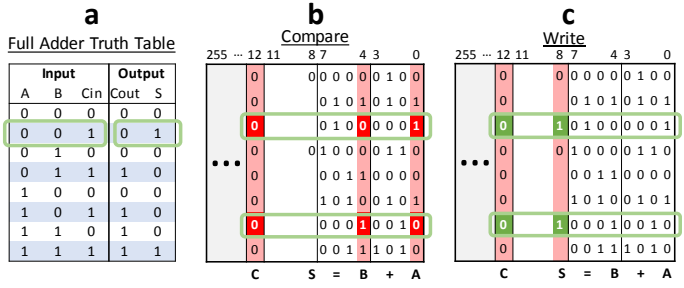

Figure 2: Example of vector addition in a CAM, for two 4-bit vectors A and B, snapshot at zero bit, 2nd entry of the truth table: (a) Full Adder Truth Table, (b) Compare, only C, $A_0$ and $B_0$ are affected, (c) Write, only C and $S_0$ in the tagged rows (PUs) are affected

During compare (Figure 2b), the input pattern '001' is compared against bit columns C, A0 and B0, for all vector elements in parallel. The matching rows (two in this example) are tagged. During write (Figure 2c), the output pattern '01' is written in bit columns C and S0 accordingly. Only the tagged rows are written. Each compare and write pair affect the entire dataset (vectors A, B and S).

### 2.2 Smith-Waterman Local Sequence Alignment Algorithm

Searching for similarities in pairs of protein and DNA sequences (also called Pairwise Alignment) has become a routine procedure in computational biology and it is a crucial operation in many bioinformatics tasks. The Smith-Waterman algorithm (S-W) [12] provides an optimal solution for the pairwise sequence alignment problem, but requires a number of operations proportional to the product of the two sequence lengths.

The Smith-Waterman algorithm identifies the local alignment of two sequences (usually DNA or protein) by computing a two-dimensional scoring matrix $H$ [12]. Each $H_{i,j}$ element is calculated according to Eq. (3). $\sigma(A_i, B_j)$ is the match score between the $i^{th}$ element of sequence $A$ and the $j^{th}$ element of sequence $B$. Matching elements score positively (e.g., +2), while mismatching results in negative score (e.g., -1). The optimal alignment score between two sequences is the highest score in the matrix $H$.

$$E_{i,j} = \max\{E_{i,j-1} - G_{ext}, H_{i,j-1} - G_{first}\} \quad (1)$$

$$F_{i,j} = \max\{F_{i-1,j} - G_{ext}, H_{i-1,j} - G_{first}\} \quad (2)$$

$$H_{i,j} = \max\{H_{i-1,j-1} + \sigma(a_i, b_i), E_{i,j}, F_{i,j}, 0\} \quad (3)$$

The alignment may contain gaps in both sequences, which are penalized in the score calculation (by negative scores). According to the affine gap model [22], opening a gap is harder than extending it, and therefore the penalty for opening a gap is larger. The affine penalty scheme is calculated with two additional matrices, $E$ and $F$, equations (1) and (2); $G_{first}$ and $G_{ext}$ are the penalties for starting and extending a gap, respectively. The matrices $E, F$ and $H$ are initialized with $E_{0,j} = E_{i,0} = F_{0,j} = F_{i,0} = H_{0,j} = H_{i,0} = 0$ for all $i$ and $j$.

Filling the scoring matrix $H$ is the computationally intensive part of S-W. In a sequential implementation of the algorithm, cells are filled in either row- or column-wise order. A parallel implementation allows all independent cells to be computed in the same iteration. Such cells reside on the same antidiagonal. The matrix is filled along the main diagonal.

#### 2.2.1 Global and Semi-Global Sequence Alignment

Global sequence alignment is a special case of local alignment. When comparing similar sequences and of roughly similar length, global alignment [23] is used, where the goal is to align every base (bp or amino acid) from each sequence. The global sequence alignment algorithm is also based on dynamic programming and it is very similar to local sequence alignment. The score of each cell is determined only from the scores of the neighboring cells, as described in eq (4):

$$H_{i,j} = \max \begin{Bmatrix} H_{i-1,j-1} + \sigma(A_{i-1}, B_{j-1}) \\ H_{i-1,j} + d \\ H_{i,j-1} + d \end{Bmatrix} \quad (4)$$

where $d$ is the gap (indel) penalty. Unlike in local alignment, the only boundary condition in global alignment is $H_{0,0} = 0$.

Semi-global alignment allows for gaps at the beginning and end of each sequence and can be viewed as a hybrid between global and local alignment. Semi-global alignment has the same boundary conditions as local alignment, but its scoring rule is identical to that of global alignment.

## 3 RESISTIVE NAND CAM ARRAY

CAM can be of either NOR or NAND type [24]. The main difference between the two is Match line organization. In NOR CAM, the Match line forms a wired NOR of all bits in the row and discharges on mismatch while retaining high voltage on match. In contrast, in NAND CAM, the Match line forms a wired NAND of all bits in the row and discharges on match, while retaining high voltage on mismatch. This work introduces a novel NAND resistive CAM (ReCAM) design, detailing its components bottom-top, from the NAND bitcell design to an associative processing array, and comparing it to an existing NOR ReCAM design [19].

### 3.1 Memristors

Resistive memories store information by modulating the resistance of nanoscale storage elements (memristors). They are nonvolatile, free of leakage power, and can be viewed as potential alternatives to charge-based memories, including NAND flash [25]. Memristors (also denoted below as R) are two-terminal devices whose resistance is changed by electrical current or voltage. The resistance of the memristor is bounded by a minimum resistance $R_{ON}$ (low resistive state) and a maximum resistance $R_{OFF}$ (high resistive state).

### 3.2 NAND ReCAM Bitcell

Figure 3a presents a 2T2R NAND ReCAM bitcell. The cell is composed of two memristors with opposing polarities and two NMOS transistors. The top transistor is used for match/mismatch evaluation. The bottom transistor is used as a write selector. Memristor resistances are used to store the bit value. Logic '1' is stored by setting the left (Bit) memristor to $R_{ON}$ and the $\overline{\text{Bit}}$ memristor to $R_{OFF}$. Logic '0' is stored by setting the Bit memristor to $R_{OFF}$ and the $\overline{\text{Bit}}$ memristor to $R_{ON}$. Each bitcell memristor serves as an ON/OFF switch for the top transistor gate (also referred to as *midpoint* in this paper), depending on the resistance. At the beginning of a compare cycle, the search pattern is set on the Bit and $\overline{\text{Bit}}$ lines ('1' on Bit and '0' on $\overline{\text{Bit}}$ to compare to '1', and '0' on Bit and '1' on $\overline{\text{Bit}}$ to compare to '0'). Figure 3b presents an example of a comparison to '1' on a NAND bitcell storing '1'. The $R_{ON}$ memristor on the left acts as a closed switch, passing a logical '1' to the gate of the top transistor, thus "opening" it. Figure 3c presents the opposite case, where the same bitcell is compared to '0'. Logic '0' is passed to the gate of the top transistor through the right ($\overline{\text{Bit}}$) memristor, keeping the transistor "closed". A bitcell is masked out by keeping both Bit and $\overline{\text{Bit}}$ lines high, ensuring that the top transistor will be open by applying '1' at its gate.

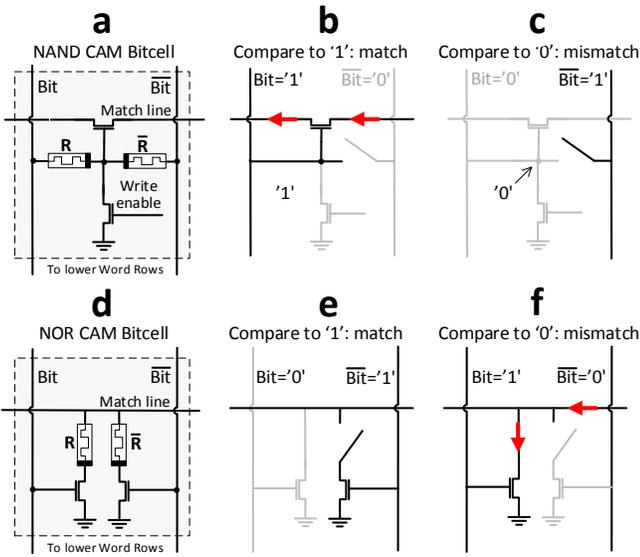

Figure 3: (a) NAND CAM 2T2R bitcell. (b), (c) Examples of a bitcell containing '1' compared to '1' and '0', allowing or blocking the match current. The left (right) memristor has $R_{ON}$ ($R_{OFF}$) resistance, depicted as an on (off) switch. (d) NOR CAM 2T2R bitcell. (e), (f) Examples of a bitcell containing '1' compared to '1' and '0', blocking or allowing the match current, opposite to NAND. Note that the compare pattern is inverted.

### 3.3 NAND Sub-Word, Word Row and Word Block

A NAND Sub-Word (Figure 4a) is composed of 32 bitcells, with a single PMOS transistor for precharge, an inverting buffer, and an evaluation switch (single NMOS transistor). Every Sub-Word receives a total of 66 control signals: 32 KEY bits times two for Bit and $\overline{\text{Bit}}$, one evaluation signal, and one precharge control (each Sub-Word has its own precharge control line, denoted as *PC(0)*, *PC(1)*, etc. in Figure 4b).

Eight Sub-Words form a Word Row (Figure 4b). In total, a Word Row contains 256 bits. All Sub-Word Match lines are ANDed to generate a match signal, which is sampled by the TAG logic at a rising clock edge. 1024 consecutive Word Rows, sharing control signal drivers, form a Word Block (Figure 4c). $N$ Word Blocks are cascaded to logically form an array of 1024$N$ rows (Figure 4d). All TAGs of consecutive Word Rows are daisy-chained to allow for inter-row data shifts (shift-down, Section 4.2).

### 3.4 Functionality

Three cycle types are defined: compare, write, and shift-down. The shift-down cycle does not involve the memory array and is explained in Section 4.2.

A compare cycle is composed of two phases, a precharge followed by an evaluation. The phases are timed by the system clock. The first phase, precharge, is enabled during the high clock phase. Every Bit and $\overline{\text{Bit}}$ line is either set according to the search pattern (KEY) or is set to '1' to mask such a bit column, precharge control is activated, and the evaluation switch is turned off (Figure 4a, left). Precharge timing is defined by the driver and Match line propagation times. During the second clock phase, evaluation, precharge is disabled and the evaluation switch is turned on, allowing the Match line to discharge in case of a match. In the case of a mismatch, at least one of the bitcell top transistors is closed, preventing the Match line discharge.

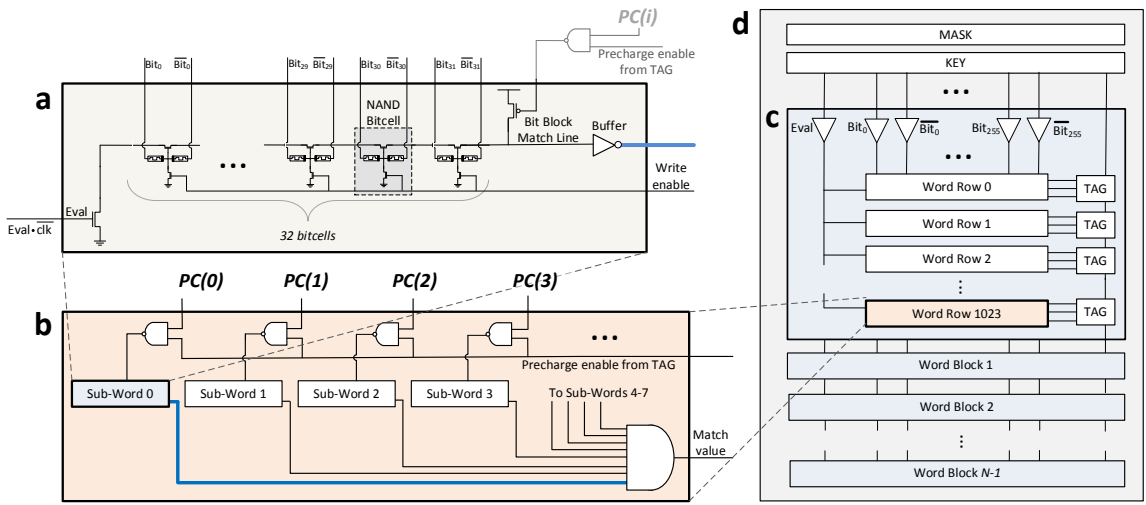

Figure 4: NAND ReCAM structure. (a) A Sub-Word structure. 32 bitcells are connected in series to an evaluation switch (NMOS transistor). (b) A Word Row structure composed of eight Sub-Words. The AND output (Match value) contains the row match/mismatch result after the evaluation phase. (c) A Word Block composed of 1024 Word Rows, sharing Bit, $\overline{\text{Bit}}$ and eval line drivers. Every Word Row has a TAG logic. (d) A NAND ReCAM chip composed of $N$ Word Blocks.

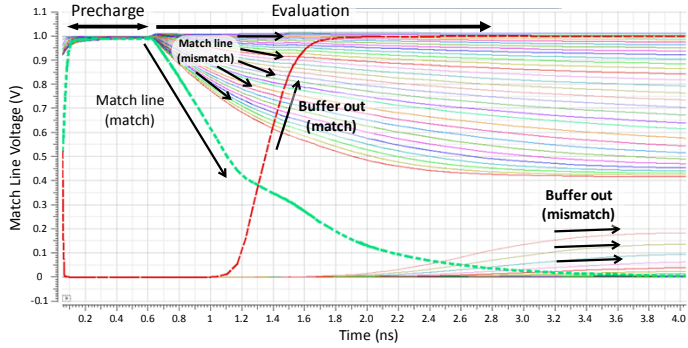

Figure 5: Spectre simulations of precharge and evaluation stages in a single Sub-Word. All mismatch cases of a single mismatching bitcell are shown. The mismatch cases typically leak down to a value above 0.4V. A matching line discharges below 0.4V within 1ns after start of evaluation, facilitating a prompt decision.

A write cycle is initiated by asserting the Write enable (Figure 4a, right), connecting all mid-points (nodes common to both transistors and both memristors) to the ground. A write operation is performed in two steps. In the first step, appropriate positive ($V_{SET}$) and negative ($-V_{SET}$) memristor setting voltages are applied only to Bit lines, setting the applicable Bit memristors to high ($R_{OFF}$) or low ($R_{ON}$) resistance, respectively. All $\overline{\text{Bit}}$ lines are kept disconnected. In the second step, the complement values are written. Voltage levels are applied only to the $\overline{\text{Bit}}$ lines, while all Bit lines are kept disconnected.

### 3.5 Implementation

A NAND Sub-Word circuit was designed and laid out using the 28nm CMOS High-k Metal Gate library from Global Foundries for transistor sizing, timing and power analysis. We performed Spectre simulations for the FF and SS corners at $70^0$c and nominal voltage. Precharge timing is at least 500ps, while evaluation is at most 1ns for mismatch (time until 90% of Vdd at buffer output). This allows operation with a 2ns cycle (500MHz). The power dissipated at precharge, when running at 500MHz, is at least 39nW (a mismatch, no Match line discharge) and at most 2.2μW (in the case of a match, followed by a Match line discharge) for a single Sub-Word. Figure 5 presents a Sub-Word Match line and buffer output voltages of a Spectre simulation for a match and all 32 cases of a single bit mismatch. We have manually laid out a $24F^2$ NAND ReCAM bitcell. The total Word Row area in 28nm technology, including the evaluation, precharge, buffer and inter Sub-Word logic, is $11\mu m^2$. Estimations for the entire chip, including the TAG logic, are presented in Section 4.3.

Memristor switching time is in the range of hundreds of picoseconds [26], while the switching energy could be as low as a few tens of femtojoules [27], which might still be prohibitively high for very large ReCAM arrays (over 1G bitcells). However, switching energy is dependent on the memristor material. As resistive memory technologies continue to evolve, we expect that future devices may achieve lower switching energy. Another factor potentially limiting the use of resistive materials in associative processing is memristor endurance. In our evaluations, we assume a memristor device with high $R_{OFF}$ [28] resistance (so that the leakage is negligible) and 100fJ memristor switching energy.

TABLE 1
AREA AND ENERGY CONSUMPTION OF NOR AND NAND SUB-WORDS. MULTIPLE COMPARES ARE PERFORMED ON EVERY WORD ROW DURING AN AP TRUTH TABLE EXECUTION, RESULTING IN A MATCH ONLY ONCE, AND IN A MISMATCH MULTIPLE TIMES: THREE (SEVEN) TIMES FOR 2-BIT (3-BIT) AP OPERATIONS

| Sub-Word Parameter | NOR Array | NAND Array | Occurrences per AP Operation |
|---|---|---|---|
| Area ($\mu m^2$) | 0.24 | 0.71 | - |
| Compare energy, match (fJ) | 2.6 | 4.4 | 1 |
| Compare energy, mismatch (fJ) | 2.8 | 1.2 (avg.) | 3 (2-bit op.) 7 (3-bit op.) |
| **Weighted Parameter** | **NOR Array** | **NAND Array** | **Difference** |
| Compare energy per **2-bit** operation (fJ) | 11 | 8 | -27.2% |
| Compare energy per **3-bit** operation (fJ) | 22.2 | 12.8 | -42.3% |

### 3.6 NAND vs. NOR ReCAM

A typical 2T2R NOR bitcell is depicted in Figure 3d. Storing a logical '1' is achieved by setting the Bit memristor to $R_{ON}$ and the $\overline{\text{Bit}}$ memristor to $R_{OFF}$. Logic '0' is stored by setting the Bit memristor to $R_{OFF}$ and the $\overline{\text{Bit}}$ memristor to $R_{ON}$. During compare, the Match line is precharged and an inverted search pattern is set on the Bit and $\overline{\text{Bit}}$ lines. Figure 3e presents a NOR bitcell with stored '1', compared against '1', resulting in a match without Match line discharge. Figure 3f shows comparison of the same bitcell against '0', resulting in a mismatch and a Match line discharge. A mismatching bitcell enables a discharge path through an $R_{ON}$ memristor from the Match line to the ground.

Table 1 presents area and compare energy of NOR and NAND Sub-Words. Area-wise, a NOR bitcell is smaller than a NAND bitcell ($8F^2$ vs. $24F^2$). In addition, a NOR Sub-Word requires a smaller buffer and does not have the evaluation transistor. Overall, a NOR Sub-Word is 2.95× smaller than a NAND.

Energy-wise, NOR match and mismatch energies are similar (2.6fJ and 2.8fJ, respectively) due to leakage and buffer static energy of a match. For NAND, the minimal mismatch energy is 0.08fJ and the maximal mismatch energy is 3.4fJ, depending on which bitcells mismatch (closest or furthest from the precharge point). Average mismatch energy is 1.2fJ.

A NAND Sub-Word does not always provide better energy efficiency. However, when it is used in a context of associative processing, where a majority of compares result in a mismatch, the precharge is saved and the discharge energy becomes significant. Therefore, the combined compare energy per operation of a NAND Sub-Word is lower than that of a NOR Sub-Word.

### 3.7 Effects of Process Variation

A number of device parameters can influence the functionality and timing of a NAND Sub-Word, which can be affected by process variation. Such parameters include the $R_{OFF}$ to $R_{ON}$ ratio, which affects the behavior of the voltage divider at the gate of the top NAND bitcell transistor; $R_{OFF}$, which affects the static power; and memristor programming time, which affects write timing. Hu et al. [29] found that the worst case $R_{OFF}$ to

TABLE 2

FULL-ADDITION TRUTH TABLE REORDERED FOR BATCH-WRITE WITH MATCHING CYCLE TYPES FOR EACH ENTRY ('CMP'=COMPARE, 'WR'=WRITE). CONSECUTIVE INPUTS WITH THE SAME OUTPUT HAVE THE SAME BACKGROUND COLOR. ON A TYPICAL AP, THE OPERATION REQUIRES 16 CYCLES. THE BATCH-WRITE OPERATION REQUIRES ONLY 12 CYCLES

| Input | | | Output | | Batch-Write Cycle Types |
|---|---|---|---|---|---|
| A | B | $C_{in}$ | $C_{out}$ | S | |
| 0 | 0 | 0 | 0 | 0 | Cmp & Wr |
| 0 | 0 | 1 | 1 | 0 | Cmp |
| 0 | 1 | 0 | 1 | 0 | Cmp |
| 1 | 0 | 0 | 1 | 0 | Cmp & Wr |
| 0 | 1 | 1 | 0 | 1 | Cmp |
| 1 | 0 | 1 | 0 | 1 | Cmp |
| 1 | 1 | 0 | 0 | 1 | Cmp & Wr |
| 1 | 1 | 1 | 1 | 1 | Cmp & Wr |

TABLE 3

ASSOCIATIVE OPERATIONS COMPOSING A FULL ADDITION SEQUENCE ('C'=COMPARE, 'W'=WRITE). HIGHLIGHTED WRITE CYCLES CAN BE BATCHED. UNDERLINED COMPARE CYCLES ARE BATCHED

| Cycle # | 1 | 2 | 3 | 4 | 5 | 6 | 7 | 8 | 9 | 10 | 11 | 12 | 13 | 14 | 15 | 16 |
|---|---|---|---|---|---|---|---|---|---|---|---|---|---|---|---|---|
| Baseline | C | W | C | W | C | W | C | W | C | W | C | W | C | W | C | W |
| With Batch-Write | C | W | <u>C</u> | <u>C</u> | <u>C</u> | W | <u>C</u> | <u>C</u> | <u>C</u> | W | C | W | | | | |

$R_{ON}$ ratio swing due to process variation is ±13%. Zangeneh et al. [30] found that the worst-case static power swing is ±6.8%. The worst-case program time and energy increase is 8.7% and 9% respectively due to voltage variation, and 17.8% and 12% respectively due to process variation. According to our simulations, while process and voltage variation do affect the timing and energy consumption of a NAND Sub-Word, as well as its endurance, its functionality remains process variation tolerant.

## 4 BATCH-WRITE ASSOCIATIVE PROCESSING

Associative arithmetic is typically bit-serial. The number of cycles required to execute a single-bit operation in associative processing is usually constant and equals twice the number of input bit combinations (first cycle for compare, second cycle for write in the matching Word Rows). For example, a naïve implementation of 1-bit full addition (three input bits) requires sixteen cycles [20]. In many arithmetic operations the same output value repeats for different inputs. Table 2 shows a reordered full addition truth table where the entries having the same output are placed consecutively. The output combinations '01' and '10' are each repeated three times. By batching multiple writes of the same value, it is possible to reduce the number of execution cycles of such an operation. In the example of Table 2, four cycles, equal to 25% of the total number of cycles, can be spared.

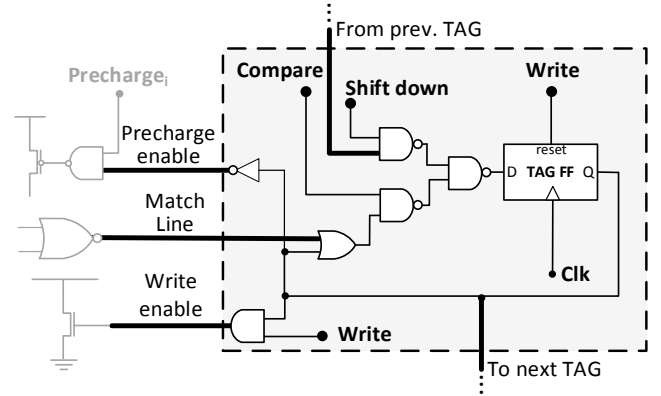

Figure 6: TAG logic supporting batch-write. Connections ending with a dot are received from the BioSEAL microcontroller.

### 4.1 Batch-Write TAG Function

In ReCAM, the TAG is responsible for storing compare results and to tag the matching row for the consecutive write. Therefore, every compare has to be followed by a write for the associative processor to function correctly. In this work, we redesign the TAG to enable a sequence of compare operations to be followed by a single write in all the matching rows. This allows the new TAG to combine the processing of all input combinations with the same output combination (for example, entries 2 through 4 and 5 through 7 in Table 2), which may lead to a significant reduction in the total number of processing cycles. We name the new TAG cell Batch-Write TAG (BW-TAG). Table 3 demonstrates a sequence of compare and write operations of a full addition, with and without batch write. Each BW-TAG accumulates the results of consecutive compares, while each write operation resets the TAGs.

Batch-write has two main advantages. The first advantage is the reduction in the number of execution cycles. For example, a 1-bit full addition, for which the truth-table is shown in Table 2, takes 16 cycles on a ReCAM without BW-TAG. On a ReCAM with BW-TAG, as demonstrated in Table 3, four write cycles are spared, leading to a 25% cycle time reduction. Table 4 summarizes several bitwise Boolean and arithmetic operations supported by ReCAM. Most associative arithmetic and logic operations are performed bit serially. Therefore, the performance improvement of a serial operation manifests in the full wordlength operation, e.g., 25% performance improvement of 1-bit full addition manifests in the same improvement for 32-bit full addition.

The second advantage of the BW-TAG is that it saves the precharge operation in ReCAM rows that were matched during the same compare sequence. The *Precharge enable* output at the top-left of Figure 6 is '0' after a match, disabling all Sub-Word precharges in a Word Row. The main benefit in this case is the reduced energy consumption.

The BW-TAG requires additional logic compared with a non-BW TAG, resulting in higher TAG energy consumption. However, the precharge disabling neutralizes this energy overhead almost entirely. In some cases, BW-TAG leads not only to performance improvement but also to reduced energy consumption. Table 4 presents performance and energy comparisons of a NAND ReCAM with and without the BW-TAG logic for specific associative operations.

TABLE 4
PERFORMANCE AND ENERGY CONSUMPTION OF NAND ReCAM WITH AND WITHOUT BW-TAG AND MAXIMAL SAVED ENERGY WITH EFFICIENT OPERAND MAPPING (EOM, SECTION 5.2.2)

| Operation | ReCAM cycles | BW-ReCAM Cycles | Perf. Impr. | BW-ReCAM Energy Overhead | EOM Energy Impact |
|---|---|---|---|---|---|
| Bitwise Boolean and Arithmetic Operations | | | | | |
| AND | 8 | 5 | 37.5% | 2.3% | -14.6% |
| OR | 8 | 5 | 37.5% | 2.3% | -14.6% |
| XOR | 8 | 6 | 25% | 3.0% | -16.7% |
| Half Add | 8 | 7 | 12.5% | 1.6% | -16.7% |
| Full Add | 16 | 12 | 25% | -0.5% | -40.9% |
| Computational Biology Operations | | | | | |
| DNA Base-pair match | 10 | 7 | 30% | -0.3% | -42.3% |
| Amino-acid match (BLOSUM62) | 1058 | 544 | 48.5% | -7.1% | -84.8% |

TABLE 5
WORD ROW AND ITS TAG ENERGY BREAKDOWN BY OPERATION TYPES. COMPARE AND WRITE ARE FOR A 1-BIT FULL ADDER (THREE INPUT BITS). MINIMAL ENERGY IS WHEN ALL INPUT BITS RESIDE IN THE SAME SUB-WORD, MAXIMAL ENERGY IS WHEN EACH INPUT BIT RESIDES IN A SEPARATE SUB-WORD

| Operation | Energy (min/max) | Sub-Words | Logic |
|---|---|---|---|
| Compare mismatch | 0.35fJ / 11fJ | 22% / 97% | 78% / 3% |
| Compare match | 10fJ / 19fJ | 43% / 70% | 57% / 30% |
| 1-bit write | 206fJ | 97.3% | 2.7% |
| 1-bit shift-down (avg.) | 217fJ | 95% | 5.3% |

## 4.2 BW-TAG Circuit Design

Figure 6 presents the BW-TAG logic circuit. The *Compare*, *Write*, and *Shift-down* signals are issued by an external microcontroller (not shown).

A sequence of consecutive compare cycles (without an intervening write) is performed as follows. For each matching Word Row, the compare result is OR–ed with the existing TAG value, thus allowing match accumulation. Once a row matches, its TAG will hold '1' until the end of the compare sequence, blocking the Match line precharge. All TAGs are reset when a write cycle is issued.

The shift-down command enables inter-PU communication, by allowing an entire ReCAM field (an arbitrary number of bit-columns) to move down by one row position. Inter-PU communication is performed bit-serially, with each bit move taking three cycles: (1) Compare with the KEY='x…x1x…x' pattern, where '1' is asserted in the bit-column being moved, while the rest of the bit columns are masked, which effectively copies the bit-column to the TAG, (2) Shift-down and (3) Write with the same KEY pattern, which effectively copies TAG to the same bit column. The input value from the previous TAG (as depicted in the upper rows of Figure 4d) is sampled, overwriting the previously stored value. Shifting down multiple data rows is possible by several consecutive shift-down cycles.

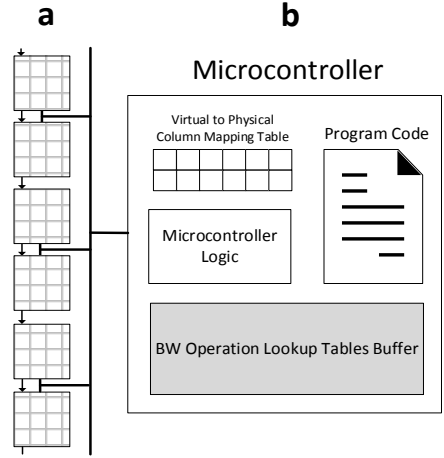

Figure 7: BioSEAL system, composed of (a) separate daisy-chained multiple NAND ReCAM modules and (b) a BioSEAL microcontroller.

During a write cycle, the *write* signal from a BioSEAL microcontroller is asserted, activating the bottom transistor in all bitcells of all tagged rows, allowing write to occur in parallel in all tagged rows. In addition, an active *write* signal invokes a TAG reset.

## 4.3 Area and Power Estimation

We synthesized the BW-TAG circuit using the Synopsis Design Compiler [31] with a Synopsis 28nm library. Following [32], we scaled the area and power figures to 28nm node technology. Estimated single BW-TAG cell area in 28nm is 7.4μm$^2$, while its power consumption at 500Mhz is 3 μW. In comparison, the non-BW TAG has a cell area of 5.3μm$^2$ and 1.9μW power dissipation (same 28nm, 500MHz operating frequency). An entire BW-ReCAM die of 230mm$^2$ can contain $2^{24}$ Word Rows, namely $2^{32}$ bits. Table 5 presents the energy consumed by a Word Row and its TAG during compare of a 1-bit full adder operation (a typical associative processing operation in DNA sequence alignment), single write and single shift-down operations.

## 5 BIOSEAL PROCESSING-IN-MEMORY SYSTEM ARCHITECTURE

Figure 7 presents a high-level view of the NAND ReCAM based BioSEAL processing-in-memory system. Daisy-chain connectivity allows an easy partitioning of BioSEAL into separate ICs (Figure 7a). The daisy chaining is only used in shift-down operations and does not limit the associative processing parallelism. The BioSEAL microcontroller (Figure 7b) is responsible for managing the processing operations by issuing instructions, setting the KEY and MASK registers, and handling control sequences. The microcontroller may also perform some baseline processing, such as normalization or aggregation of data from different ICs.

The scaling of BioSEAL is relatively simple. The inherent parallelism of BioSEAL allows an almost linear increase in performance by cascading BioSEAL ICs as the biological datasets grow along with the memory footprint. Since the bulk of data is never transferred outside the memory arrays through a bandwidth-limited communication interface, the bandwidth-wall performance limit suffered by von Neumann architecture is pushed further away.

## 5.1 Programming BioSEAL

An external host may run an operating system and sequential code, and may delegate sequence alignment tasks to BioSEAL. The code intended to run in BioSEAL is translated into associative processing primitives that are downloaded into and executed by the BioSEAL microcontroller. Presently, BioSEAL code is manually encoded at assembly language level.

The host invokes BioSEAL to perform its code fraction. BioSEAL receives the workload parameters and starts execution. Once BioSEAL execution completes, the host can access the BioSEAL output. There is no hardware support for data coherence between the host CPU and BioSEAL. BioSEAL has no access to the host main memory or on-chip cache. Therefore, the datasets on which BioSEAL operates must reside in BioSEAL and should not be left in the host memory. To avoid inconsistencies between the BioSEAL and host CPU memory, BioSEAL memory is inaccessible to the host CPU during BioSEAL operation.

## 5.2 System-Level Optimizations

Two optimizations are employed by the BioSEAL microcontroller to improve the system energy efficiency. The first one blocks precharge in masked-out Sub-Words. The other method optimally allocates data operands within the BW-ReCAM array.

### 5.2.1 Disabling Precharge of Masked-Out Sub-Words

Each Sub-Word in a Word Row receives a separate precharge control signal from the microcontroller. When the microcontroller identifies that a certain Sub-Word is masked-out, the precharge control disables the precharge in such Sub-Word. Evaluation is still performed in the masked Sub-Words. As a result, masked Sub-Words keep the Match line low, as if a match has occurred. The goal of this approach is to prevent, during execution time, the precharge in as many Sub-Words as possible. In a typical associative bit-serial arithmetic or logic operation, only two or three bit-columns are active (*cf.* Table 2). In these cases, at most three Sub-Words are active and should be precharged. The rest are masked and the precharge energy is saved. For example, in a full addition, the highest energy case is when the three input bits reside in separate Sub-Words. In that case, the compare energy reduction is 2.5× (from eight active Sub-Words to three). If only one Sub-Word is active during compare, 6.7× compare energy is reduced.

### 5.2.2 Efficient Operand Mapping (EOM)

The second optimization allocates at compile time the operands of an operation (e.g., operands A and B of an accumulation operation B←A+B) in the same Sub-Word. With precharge blocked, as proposed in the previous section, the masked Sub-Words do not consume dynamic power. Therefore, our goal is to minimize the number of active Sub-Words during a compare. In each operation, not all operand bits have to reside in the same Sub-Word. Since most operations are performed bit-serially (e.g., a 32-bit integer add is decomposed into a single half-add and 31 full-add operations), it is only required that the same-addition bits from both operands be located in the same Sub-Word. Operands can be distributed across all eight Sub-Words while still maintaining the minimal number of active Sub-Words per operation. To achieve this, a column mapping table is managed by the microcontroller. The table maps between physical on-chip columns and user-assigned virtual columns for each vector variable. A pre-processing step analyses the assembly-level program and identifies pairs of operands that should reside in the same Sub-Word. The pre-processor then maps the operand bits accordingly.

Table 4 presents the energy impact of EOM optimization. Note that negative figures mean energy reduction relative to the baseline. For logic and arithmetic operations, the number of input bits is either 2 or 3, at most requiring a Sub-Word per bit. In the case of a DNA base pair or amino acid match, four or all eight Sub-Words in Word Row may be precharged. With EOM optimization, only one Sub-Word is precharged during such a match.

## 6 ENTIRE DATABASE SEQUENCE ALIGNMENT WITH BIOSEAL

S-W identifies the optimal alignment of two sequences by computing a two-dimensional scoring matrix. Matching base pairs score positively, while mismatching base pairs result in a negative score. The optimal alignment score of two sequences is the highest score in the matrix. The S-W algorithm has two steps: scoring (to find the maximal alignment score) and trace-back (to construct the alignment). Scoring is the most computationally demanding, while trace-back requires significantly less computing resources and can therefore be performed by an external host CPU following the scoring performed by BioSEAL. The scoring step is the focus of this work.

### 6.1 Pairwise Local Sequence Alignment on BioSEAL

In a parallel implementation of S-W, the matrix is filled along the main diagonal (Figure 8a) and all the antidiagonal scores are calculated in parallel. Figure 8b and c show a snapshot of two BioSEAL ReCAM ICs at the beginning (b) and end (c) of a single iteration of the algorithm. Two antidiagonals are required to calculate the score of a new anti-diagonal. Therefore, in each iteration only three anti-diagonals are stored in ReCAM. The dataset may be distributed over a large number of BioSEAL modules, possibly located in separate ICs, as presented in Figure 7.

The dynamic programming scoring matrix antidiagonals are mapped to the ReCAM array columns. The antidiagonals are stored in $AD[0]$-$[2]$ and are cyclically buffered so that the most recent two are used to compute the score of a new antidiagonal. Vectors $E$ and $F$ store partial results according to the affine gap model [22] (essentially equivalent to two antidiagonals which store partial results). $A$ and $B$ hold the two aligned sequences, one element from each per ReCAM row. A *tmp* vector is used to store temporary intermediate results. Overall, every ReCAM row retains one element of the vectors $A$, $B$, $E$, $F$, *AD[0], AD[1], AD[2]* and *tmp*.

The number of iterations equals the number of antidiagonals ($n+m-1$). At the beginning of an iteration, $B$ is shifted one row down to align the appropriate symbols of the two sequences. Figure 8b shows the BioSEAL ReCAM memory map at the beginning of an iteration. $A$ and $B$ contain the sequences and reside in a separate ReCAM row. Shift-down operations move the sequence down along the ReCAM array. Figure 8c shows the ReCAM map at the end of the iteration.

The sequential time complexity of computing all matrix scores is $O(nm)$, where $n$ and $m$ are the respective lengths of the sequences. Parallel time complexity on $p$ parallel processing units is $O(nm/p)$. In BioSEAL, the processing unit is a memory row. Since BioSEAL may comprise hundreds of millions of rows, unlike GPU or FPGA implementations, $p$ could possibly be larger than $max\{n,m\}$ even for very large $n$ and $m$. Hence, BioSEAL can achieve linear time complexity of $O(max\{n,m\})$.

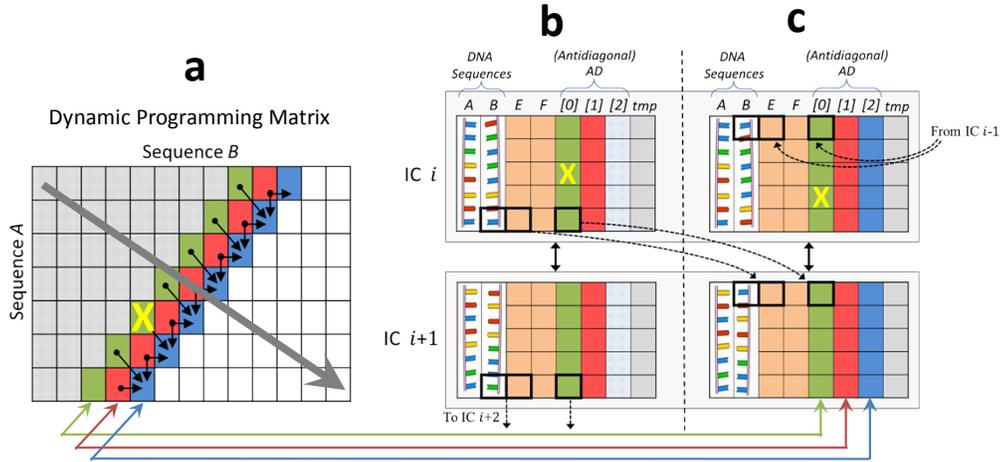

Figure 8: (a) A snapshot of the S-W dynamic programming matrix shows the direction of progress for the parallel algorithm. An example of organization of data in the BioSEAL ReCAM IC arrays at the beginning (b) and end (c) of an iteration. AD[2] contents in (b) are replaced with the new result (c).

Global and semi-global alignments use a subset of the operations required from local alignment; both can be implemented on BioSEAL with only few modifications to the local alignment implementation.

The S-W algorithm is not memory bound but compute, bound [33][34]. Hence, its performance does not necessarily gain from an in-memory implementation (although its power efficiency does, as we show in Section 7). The main sources of BioSEAL speedup are its massive fine-grain parallelism and associative processing, the combination of which enables a high performance, energy efficient implementation.

### 6.2 Multiple Parallel Pairwise Local Sequence Alignments

In [19], a pairwise in-memory associative alignment algorithm was presented. This section presents an extension of that algorithm, which aligns only single pairs of sequences; that is, it exploits only the intra-pair alignment parallelism. However, aligning a single query sequence against multiple reference sequences can be further parallelized on the associative in-memory accelerator. Alignment of the entire database is possible by applying a number of novel associative processing techniques, as presented below. This subsection introduces a novel algorithm for a multiple-pairwise alignment on BioSEAL ReCAM, while Section 6.3 shows how the algorithm can be used for entire database pairwise alignment search on BioSEAL.

A query sequence can be aligned against multiple reference sequences (sequence database search) in parallel on BioSEAL by applying an associative signaling technique as follows. We add two special-purpose flags. The first two-bit flag is stored in the two leftmost bit columns of the ReCAM. Its first bit marks the first row of a database sequence ("first_row") and its second bit marks the buffer row at the end of such sequence ("buffer_row"), which separates two consecutively stored sequences. The second two-bit flag controls the algorithm execution. Its first bit ("wavefront_rows") marks the rows where the computation is performed (added as a condition in the compare operations). The second bit ("shift_down_rows") marks the rows to be shifted during execution. Figure 9 presents a snapshot of the entire database alignment execution.

Figure 10 presents the pseudocode of the multiple pairwise alignment algorithm. The algorithm has two parts, alignment and max score reduction. The Alignment step calculates all pairwise alignment scores. Following the alignment step, all scores are stored as a vector (presented as the rightmost column in Figure 10). The max score reduction (lines 21-28 in Figure 10), computes the alignment score of each alignment. The maximal score is then stored in the buffer row.

All database sequences are stored in BioSEAL before the execution begins. During the first iteration, the first symbol (DNA base pair or protein amino-acid) of the query sequence is written to the first row of each database sequence in the query sequence columns. From the second iteration and on, the query sequence is shifted one row down. Until the query sequence is

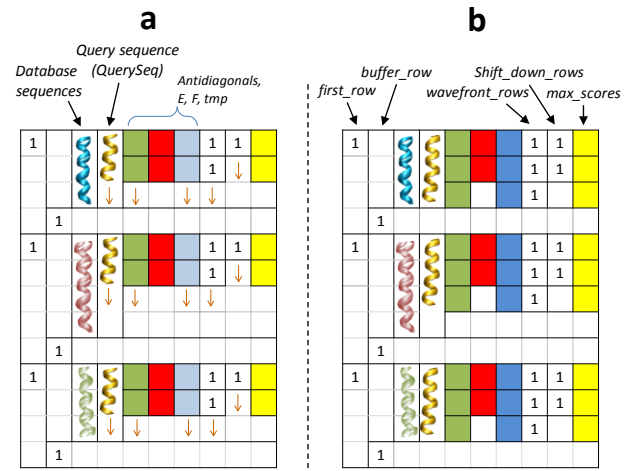

Figure 9: Snapshot of a single iteration of multiple pairwise alignment against the backdrop of a BioSEAL ReCAM map. Database sequences (blue, red, green coils) are stored in place before the execution starts. (a) Shows the beginning of an iteration while (b) shows the completion of an iteration. During the iteration, the query sequence (yellow coil) is shifted down one row, allowing the new scores to be calculated.

**Algorithm 1** Entire Database Sequence Alignment on BioSEAL

**Input**: a query sequence (*query_seq*) of length m.
**Output**: rows containing maximal alignment score of each alignment.

FINDMAXALIGNMENTSCORE(*query_seq*[], m):

   //Initialization stage
1: **copy** *wavefront_rows* **to** *shift_down_rows*

   //wavefront rows of prev. iteration are now the $shift\_down\_rows$
2: **shift_down** *active_row* **by** 1 row
3: **shift_down** $query\_seq$ **by** 1 row
4: **if** (*iteration_num* < $m$):
5:    **write** *query_seq[iteration_num]* **to** *query_seq* **where** first_row=='1'

   //Perform the pairwise alignment of [19]:
6: **for** i=0 to max_database_sequence_length+m-1 **do** {
7:    right_AD ← i mod 3; middle_AD ← (i – 1) mod 3;
      left_AD ← (i – 2) mod 3
8:    **shift_down** *query_seq* **where** *wavefront_rows*=='1'

   //All following instructions are performed with column conditions:
      *wavefront_rows*=='1' && *buffer_rows*=='1'
9:    **shift_down** AD[left_AD]
10:   AD[right_AD] ← AD[left_AD] + **match**(seqA, QuerySeq)
11:   AD[right_AD] ← **max**{AD[left_AD], 0}
12:   AD[left_AD] ← AD[middle_AD] - $G_{first}$
13:   tmp ← F - $G_{ext}$
14:   F ← **max**{AD[left_AD], tmp}
15:   AD[right_AD] ← **max**{AD[right_AD], F}
16:   tmp ← E - $G_{ext}$
17:   E ← **max**{AD[left_AD], tmp}
18:   **shift_down** E **where** shift_down_rows=='1'
19:   AD[right_AD] ← **max**{AD[right_AD], E}
20:   max_scores ← **max**{AD[right_AD], max_scores}
   }

   // All maximal wavefront scores are now stored in $max\_scores$ column.
   Next part reduces all scores to a single value, one for each pairwise alignment.

MAXSCOREREDUCTION:

21:   **write** '1' **in** wavefront_rows column **where** first_row=='1'
22:   **copy** tmp ← max_scores
23:   **shift_down** tmp **where** wavefront_rows=='1'
24:   **for** i=0 to $max\_database\_sequence\_length$:
25:      tmp ← **max**{tmp, max_scores}

      // calculate new $max\_scores$ for each cell
26:      **shift_down** tmp **where** wavefront_rows=='1'
27:      **shift_down** wavefront_rows
28:      **shift_down** tmp **where** wavefront_rows=='1'

   // buffer_row of each sequence now contains the maximal alignment score fully written, the next query symbol is written to the first row (similar to the first iteration). During an iteration, most of the instructions from the pairwise alignment of [19] are unchanged except for a condition to operate only on rows with 'wavefront_bit==1' or 'shift_down_bit==1', as explained in Figure 10.

Figure 10: BioSEAL multiple pairwise sequence alignment pseudocode.

The main difference between the algorithm from [19] presented in Section 6.1 and the one presented here is that the former implemented only intra-pair parallelism, while the latter implemented both intra- and inter-pair parallelism.

### 6.3 BioSEAL as a Biological Sequence Database Storage and Search

An entire database of proteins or DNA sequences can be stored and aligned in BioSEAL. The database is populated (initialized) by a combination of shift-down and write. A one-bit flag is shifted down, followed by a write of a single sequence symbol to a Word Row. Initialization is performed serially on each Word Row, and therefore its time is determined by the populated sequence length. For DNA, the longest sequences can span over several ICs (Section 7). The size of each IC, as well as the mapping of each populated symbol to every IC Word Rows, is known before the initialization. Therefore, to avoid serial initialization of one IC after another, the shift-write operation is performed in parallel on all ICs, followed by multiple writes, one for each IC symbol. The same holds true for a database of proteins or short DNA sequences.

For the alignment execution, a single ReCAM Word Row serves as a processing unit (PU). The highest parallelism is achieved when a single symbol is stored and processed by a single PU (Word Row). Therefore, specific columns are allocated in all the BioSEAL ICs to store the database sequences and their bit flags. A database search encompasses all BioSEAL chips, in parallel, with the final score of each alignment residing in the buffer row. A single final instruction to tag the highest scored pairwise alignment is then executed. Once the execution is complete, the tagged row contains the highest score in a predefined location. The buffer row can also contain metadata regarding each related database sequence, such as a sequence index, length, etc. The final score tagging can then be repeated for the second largest score and so on.

The array following a query execution is reset by resetting the TAGs (e.g., with the *write* signal) and then writing zeros to all the bit columns that store partial alignment results.

## 7 PERFORMANCE AND ENERGY EFFICIENCY COMPARISONS

Table 7 shows the performance and energy efficiency of BioSEAL as compared to a number of reference platforms that perform large-scale DNA or protein sequence alignment and database search tasks.

For DNA sequence alignment, we compare BioSEAL with SWAPHI-LS [34], RIVYERA [35], CUDAlign 4.0 [36] and PRINS [19]. SWAPHI-LS [34] is a multi-Xeon phi implementation of S-W for long sequence pairs, implementing only the scoring step of the algorithm. Six genomes from the publicly available NCBI Nucleotide database, with lengths varying from 4.4M base pairs (bps) up to 50M bps (denoted D50M in Table 7), were used for evaluations. The performance was evaluated on a compute node with two Intel E5-2670 8-core CPUs, 64GB of RAM and 4 Xeon Phis (each running at 1.05GHz with 7.9GB device memory). Performance figures were reported in [34] while power figures were taken from the manufacturer spec sheets [37], [38]

In RIVYERA [35], a S-W implementation on 128 Xilinx S6-LX150 FPGAs is presented. The dataset contains 1 million short reads of 100bps each, aligned against an entire human genome (hg19, approx. 3B bps), focusing only on the scoring step of the algorithm. On BioSEAL, this task is implemented as a database sequence search using the algorithm presented in Section 6.2, with the short reads acting as the database sequences and the human genome as the query sequence. In this case, a 9-bit integer operation is sufficient for alignment. In general, associative processors natively support variable wordlength. Both performance and power figures of the entire system were reported [35].

TABLE 7
PERFORMANCE AND ENERGY EFFICIENCY COMPARISONS OF BioSEAL WITH OTHER WORKS ('Q'=QUERY SEQUENCE)

| Compared Work Ref. | Architecture | Dataset | Achieved perf. (TCUPS) | BioSEAL performance (TCUPS) | Energy efficiency (GCUPS/W) | BioSEAL energy efficiency (GCUPS/W) | Required BioSEAL ReCAM chips |
|---|---|---|---|---|---|---|---|
| **DNA Sequence Alignment** | | | | | | | |
| SWAPHI-LS [34] | 2 CPUs 4 Xeon Phis | D33M vs. D42M | 0.133 | 5 | 0.17 | 15.8 | 3 |
| | | D33M vs. D50M | 0.106 | 5.4 | 0.13 | 16 | 3 |
| | | D42M vs. D50M | 0.111 | 6.3 | 0.14 | 16.3 | 3 |
| RIVIYERA [35] | 128 Xilinx FPGAs | hg19 vs. 1M 100bp reads | 6 | 90.7 | 7.7 | 53 | 15 |
| CUDAlign 4.0 [36] | 384 Tesla M2090s | Chr. 5 (180M×183M) | 11.1 | 25.1 | 0.111 | 16.3 | 12 |
| | | Chr. 8 (146M×144M) | 10.4 | 20.4 | 0.104 | 16.2 | 10 |
| PRINS [19] | ReCAM | Chr. 1 (249M×228M) | 27.3 | 32.8 | 15.1 | 16.4 | 15 |
| | | Chr. 5 (180M×183M) | 20.9 | 24.2 | 15.0 | 16.3 | 12 |
| | | Chr. 8 (146M×144M) | 17 | 19.7 | 14.9 | 16.2 | 10 |
| **Protein Sequence Alignment** | | | | | | | |
| SWhybrid [41] | Dual E5-2683v4 1 Titan X 1 GTX 1080 | Q: P0C6B8 (len=3564) | 1.03 | 4.8 | 1.54 | 13.6 | 12 |
| | | Q: P08519 (len=4548) | 1.05 | 6 | 1.57 | 15.2 | 12 |
| | | Q: Q9UKN1 (len=5335) | 1.03 | 6.9 | 1.54 | 16.2 | 12 |
| PRINS [19] | ReCAM | Q: P0C6B8 (len=3564) | 3.4 | 4.8 | 9.8 | 13.6 | 12 |
| | | Q: P08519 (len=4548) | 4.2 | 6 | 11.3 | 15.2 | 12 |
| | | Q: Q9UKN1 (len=5335) | 4.8 | 6.9 | 12.3 | 16.2 | 12 |

TABLE 6
NUMBER OF CYCLES OF EACH TYPE, BELONGING TO A SINGLE ITERATION OF EACH ALGORITHM

| Algorithm | # of Compares | # of writes | # of Shift-Downs | Max Chip Power |
|---|---|---|---|---|
| DNA sequence alignment | 797 | 419 | 66 | 262W |
| DNA database search (reduction) | 96 | 48 | 33 | 226W |
| Protein sequence alignment | 963 | 263 | 42 | 187W |
| Protein database search (reduction) | 60 | 30 | 24 | 226W |

CUDAlign 4.0 [36] is a multi-GPU framework for executing very large sequence alignment tasks. Evaluations were performed on the XSEDE Keenland Full Scale system with 128 nodes, each with two 8-core Intel Sandy Bridge CPUs, three NVIDIA M2090 GPUs and 32GB or RAM. The dataset was all human (GRCh37) and chimpanzee (panTro4) homologous chromosomes, taken from the NCBI database. Matching pairs of chromosome sequences were aligned (lengths denoted in Table 7). CUDAlign 4.0 [36] returns the complete alignment of two sequences. The performance of every execution stage, including that of the scoring stage alone (stage 1 [36]), was presented. As reported [36], the overhead on the scoring step required to support traceback is less than 1% of the entire scoring duration. CUDAlign 4.0 power figures were taken from the processing elements manufacturer spec sheets [39], [40].

In PRINS [19], a NOR ReCAM processing-in-storage architecture was presented, with performance and energy efficiency results on the same dataset as [36] with the addition of chromosome 1, the longest human chromosome. For an apples-to-apples comparison of performance and energy efficiency of PRINS and BioSEAL, we performed the same synthesis and simulations as described in Sections 3.5 and 4.3 on both PRINS and BioSEAL.

For protein database search we compare with PRINS [19] and SWhybrid [41], a framework for protein sequence database search on heterogeneous architectures. Evaluation was performed on several query sequences on the UniProtKB Swiss-Prot protein database [42], which contains 550k sequences with 200M amino acids. Query sequences were also taken from the Swiss-Prot database.

SWhybrid was evaluated on several platforms in [41], where only the scoring step was considered. We compare with the highest-performing of these platforms, which contain dual Intel E5-2683v4 CPUs, a Maxwell Titan X GPU, and a Pascal GTX 1080 GPU (all four units participate in the query execution). The scoring scheme used is BLOSUM62 [43]. Performance figures were taken from [41], while power figures were taken from the processing elements manufacturer spec sheets [44][45][46]. On BioSEAL, the entire Swiss-Prot database requires at most 12 chips. Using fewer ICs is possible when storing multiple sequences in a BW-ReCAM row. The tradeoff is lower parallelism.

On BioSEAL, a single iteration of DNA pairwise alignment takes 1880 cycles, while the multi-pairwise alignment execution adds 330 cycles for max score reduction iteration and bit flag handling. Table 6 presents the peak power consumption of the pairwise and multiple pairwise sequence alignment algorithms, for both DNA and protein sequences. We evaluated the performance of BioSEAL using an in-house instruction-level simulator. In accordance with simulation results and memristor parameters, a 2ns clock cycle was chosen, enabling an operational frequency of 500MHz. Energy measurements were taken from the Spectre and Design Compiler simulations reported in Sections 3.5 and 4.3. We implemented both optimizations for energy reduction presented in Section 5.2. Inter-die communication was assumed to be 1nJ per bit [47] with a 500Mbit/sec bandwidth per die. For DNA sequence alignment, sequence initialization time was taken into account. For protein, BioSEAL acts as a database that serves multiple queries; therefore, only alignment times are reported, as in [41].

Performance of the S-W algorithm was measured in Cell Updates per Second (CUPS). On the examined large scale DNA sequence alignment tasks, BioSEAL outperformed (was more energy efficient than) SWAPHI-LS by 37-57× (93-123×), CUDAlign 4.0 by 2-2.3× (146-156×) and PRINS by 1.2× (1.1×). For sequence database search, BioSEAL outperformed (was more energy efficient than) RIVYERA by 15× (7×), SWhybrid by 4.7-6.7× (8.8-10.5×) and PRINS by 1.4× (1.3-1.4×).

Compared to PRINS, BioSEAL offers a major improvement in both performance and energy for protein database search, with combined performance-power improvement of 1.9×.

## 8 RELATED WORK

### 8.1 Processing-in-Resistive Memory

Guo *et al.* used STT-MRAM and Resistive Ternary CAM for data-intensive computing [48][49]. The associative capabilities of CAM and Ternary CAM were used mainly for search operations, while the computing stayed in a host CPU. Somnath *et al.* [50] developed MBARC, a resistive crossbar in-memory LUT-based processing architecture. Chi *et al.* [51] introduced PRIME, a PIM accelerator of neural network applications. Yavits *et al.* [20] presented ReAP, a resistive CAM based massively parallel associative accelerator. In [52], a PRINS architecture was applied to data-intensive machine learning algorithms such as K-means and K-nearest neighbors. Morad *et al.* [53] introduced Re-GP-SIMD, a two-dimensional resistive processing-in-memory accelerator. Shafiee *et al.* [54] developed ISAAC, an in-situ accelerator of neural networks, where memristor crossbar arrays are used to perform dot-product operations in an analog manner. Kvatinsky *et al.* [55] and Talati *et al.* [56] developed MAGIC, a RRAM based in-data processing-in-memory architecture. Hamdioui *et al.* [57] introduced CIM, a memristor based computation in memory architecture. Bojnordi and Ipek [58] introduced the Memristive Boltzmann machine, a hardware accelerator for combinatorial optimization and deep learning. Imani *et al.* [59] introduced APIM, an approximate processing in-memory architecture which exploits the analog characteristics of resistive memory to support addition and multiplication inside the crossbar. In another work, Imani *et al.* [60] also introduced A-HAM, an analog hyperdimensional associative memory, targeted for similarity search applications. S. Khatamifard *et al.* [61] introduced a BioCAM, a resistive near memory DNA read mapping filtering accelerator. Zha *et al.* [62] introduced IMEC, a morphable resistive memory based fabric capable of supporting logic, ternary CAM, memory and interconnect functionality.

### 8.2 Sequence Alignment Accelerators

Sequence alignment acceleration solutions that implement edit-distance approximation have been proposed. Fujiki *et al.* [63] presented GenAx, an automata-based DNA sequence alignment accelerator for short reads. GenAx implements a seed-and-extend algorithm, which maps short consecutive strings called seeds (e.g., 11 bps) on the reference sequence (indexed in a hash table), and then aligns extended regions using S-W. Seed-and-extend techniques may output sub-optimal alignment results when the number of edits is larger than a certain threshold [64]. Turakhia et al. presented Darwin [65], a filtering and alignment accelerator based on two novel algorithms, that achieves high accuracy alignment results compared to existing tools on third generation long reads. Huangfu *et al.* [66] presented RADAR, a ReRAM based accelerator for BLAST that mitigates data movements by performing in-memory string comparison operations. GenAx Darwin and RADAR have implemented approximate algorithms for the local sequence alignment problem and are limited to sequences of read lengths (hundreds of bps in case of short reads, tens of thousands in case of long reads).

Reconfigurable hardware solutions have also been proposed. Ahmed *et al.* [67] presented an accelerated version of BWA-MEM using a hardware-software co-design. The suffix array lookup and seed extensions kernels of BWA-MEM were accelerated using four Xilinx Virtex-6 FPGAs. Arram *et al.* [68] presented an alignment accelerator based on the FM-index algorithm realized on a Maxeler [69] board. Banerjee [70] presented an edit-distance approximation which encodes gap penalties with circuit delays implemented in a Xilinx Virtex-7 FPGA.

Other solutions used the IBM Cell/BE architecture to implement a multi-threaded vectorized implementation of Smith-Waterman [71][72] and presented Cell implementations of a time efficient alignment algorithm for long sequences [73]. Gálvez et al. [74] presented an implementation of global sequence alignment on a Tilera Tile64 card, a multi-core architecture with 64 cores. In a later work, Gálvez et al. [75] presented a protein database search solution based on a heterogeneous architecture that used a CPU, GPU and a Xeon Phi with a peak throughput of about 140 GCUPS. Ramirez et al. [76] presented a Smith-Waterman implementation on the SARC architecture, where a heterogeneous set of processors is managed at runtime in a master-worker mode. Sarkar et al. [77] proposed Network-on-Chip architectures of sequence alignment operations and showed significant improvement over other architectures such as Cell/BE, FPGA, GPU and CPU.

## 9 CONCLUSIONS

We presented BioSEAL, a novel massively parallel processing-in-memory accelerator for large-scale DNA and protein sequence alignment tasks, including sequence database search. BioSEAL facilitates associative processing and consists of multiple ReCAM dies, conceptually serving as a large-scale associative processing array. Each BioSEAL ReCAM die can contain millions of rows, each serving as a processing unit. The BioSEAL ReCAM is based on a novel energy-efficient resistive NAND array and uses batch write to reduce the number of cycles required for associative processing operations. BioSEAL ReCAM was synthesized using 28nm technology.

We presented a database search algorithm on BioSEAL that allows an entire database of sequences to be aligned against a query sequence. We evaluated BioSEAL against several large genome sequence and protein database accelerators. Our results show up to 57× better performance and up to 156× better energy efficiency.

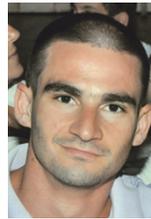
**Roman Kaplan** received his BSC and MSc from the faculty of Electrical Engineering, Technion, Israel, in 2009 and 2015, respectively. He is now a PhD candidate in the same faculty under the supervision of Prof. Ran Ginosar. Kaplan's research interests are parallel computer architectures, in-data processing, accelerators for bioinformatics, computational biology and machine learning.

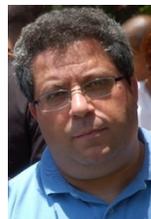
**Leonid Yavits** eceived his MSc and PhD in Electrical Engineering from the Technion. After graduating, he co-founded VisionTech where he co-designed a single chip MPEG2 codec. Following VisionTech's acquisition by Broadcom, he co-founded Horizon Semiconductors where he co-designed a Set Top Box on chip for cable and satellite TV. Leonid is a postdoc fellow in Electrical Engineering in the Technion. His research interests include non von Neumann computer architectures and processing in memory.

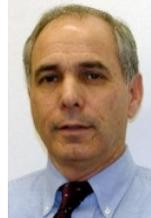
**Ran Ginosar** received his BSc from the Technion—Israel Institute of Technology in 1978 (summa cum laude) and his PhD from Princeton University, USA, in 1982, both in Electrical and Computer Engineering. His Ph.D. research focused on shared-memory multiprocessors. He joined the Technion faculty in 1983. He was a visiting Associate Professor with the University of Utah in 1989-1990, and a visiting faculty with Intel Research Labs in 1997-1999. He is a Professor at the Department of Electrical Engineering and serves as Head of the VLSI Systems Research Center at the Technion. His research interests include VLSI architecture, manycore computers, asynchronous logic and synchronization, networks on chip and biologic implant chips. He has co-founded several companies in various areas of VLSI systems.